\RequirePackage{fix-cm}
\documentclass[natbib,smallextended]{svjour3}       
\smartqed  
\usepackage{graphicx,verbatim}
\usepackage{epstopdf} 
\usepackage[colorlinks,urlcolor=blue,citecolor=blue,linkcolor=blue]{hyperref}
\usepackage{amssymb,amsmath}

 \journalname{my journal}
\newcommand\ompt{\omega_{\rm pe}t}
\newcommand{\unit}[1]{\nobreak{\mathrm{\;#1}}} 
\newcommand{\ex}[1]{10^{-#1}}

\newcommand{\be}{\begin{eqnarray}}
\newcommand{\ee}{\end{eqnarray}}
\newcommand{\bi}{\begin{itemize}}
\newcommand{\ei}{\end{itemize}}

\newcommand{\fig}[1]{Fig.~\ref{fig:#1}}

\newcommand{\ssr}{{Space Sci. Rev.}}
\newcommand{\aap}{{Astron. Astrophys.}}
\newcommand{\mnras}{{Mon. Not. R. Astron. Soc.}}
\newcommand{\pre}{{Phys. Rev. E}}
\newcommand{\prd}{{Phys. Rev. D}}
\newcommand{\apj}{{Astrophys. J.}}

\newcommand{\apjl}{{Astrophys. J. Lett.}}

\newcommand{\apjs}{{Astrophys. J. Supp.}}
\newcommand{\aapr}{{Astron. Astrophys. Rev.}}

\newcommand{\ie}{\emph{i.e.} }
\newcommand{\eg}{\emph{e.g.} }
\newcommand{\myD}{\mathcal{D}}
\newcommand{\ave}[1]{\left\langle #1\right\rangle}
\newcommand{\myp}{p}
\newcommand{\mysp}{{s_p}}
\newcommand{\ompe}{{\omega_{\rm pe}}}

\newcommand{\RecentA}[1]{#1}

\newcommand{\PNGfigure}[1]{#1}

\begin{document}

\title{Relativistic Shocks: Particle Acceleration and Magnetization
\thanks{All authors contributed equally to this review.}
}

\titlerunning{Relativistic Shocks: Particle Acceleration and Magnetization}        

\author{L.~Sironi \and U.~Keshet  \and M.~Lemoine}

\institute{
L. Sironi \at Harvard-Smithsonian Center for Astrophysics, Cambridge, MA, 02138, USA\\ \email{lsironi@cfa.harvard.edu}
\and
U. Keshet \at Physics Department, Ben-Gurion University of the Negev, Be'er-Sheva 84105, Israel \\
\email{ukeshet@bgu.ac.il}
\and
M. Lemoine \at \RecentA{Institut d'Astrophysique de Paris, CNRS - UPMC, 98 bis boulevard Arago, F-75014 Paris, France} \\
\email{lemoine@iap.fr}
}

\date{Received: date / Accepted: date}

\maketitle

\begin{abstract}
We review the physics of relativistic shocks, which are often invoked as the sources of non-thermal particles in pulsar wind nebulae (PWNe), gamma-ray bursts (GRBs), and active galactic nuclei (AGN) jets, and as possible sources of ultra-high energy cosmic-rays.
We focus on particle acceleration and magnetic field generation, and describe the recent progress in the field driven by theory advances and by the rapid development of particle-in-cell (PIC) \RecentA{simulations}.
In weakly magnetized or quasi parallel-shocks (\ie where the magnetic field is nearly aligned with the flow), particle acceleration is efficient. The accelerated particles stream ahead of the shock, where they generate strong magnetic waves which in turn scatter  the particles back \RecentA{and forth across} the shock, mediating their acceleration. In contrast, in strongly magnetized quasi-perpendicular shocks, the efficiencies of both particle acceleration and magnetic field generation are suppressed.
Particle acceleration, when efficient, modifies the turbulence around the shock on a long time scale, and the accelerated particles have a characteristic energy spectral index  of $s_\gamma\simeq 2.2$ in the ultra-relativistic limit.
We discuss how this novel understanding of particle acceleration and magnetic field generation in relativistic shocks can be applied to high-energy astrophysical phenomena, with an emphasis on PWNe and GRB afterglows.
\end{abstract}

\section{Introduction}\label{intro}
In pulsar wind nebulae (PWNe), gamma-ray bursts (GRBs), and jets from active galactic nuclei (AGNs), signatures of non-thermal processes are revealed by power-law radiation spectra spanning an extremely wide range of wavelengths, from  radio to X-rays, and beyond. Yet, it is still a mystery how the emitting particles can be accelerated up to ultra-\RecentA{relativistic} energies \RecentA{and how the strong magnetic fields are generated, as} required in order to explain the observations. In most models, non-thermal particles and near-equipartition fields are thought to be produced at relativistic shock fronts, but the details of the mechanisms of particle acceleration and magnetic field generation are still not well understood.

Particle acceleration in shocks is usually attributed to the Fermi process, where particles are energized by bouncing back and forth across the shock. Despite its importance, the Fermi process is still not understood from first principles. The highly nonlinear coupling between accelerated particles and magnetic turbulence -- which is generated by the particles, and at the same time governs their acceleration -- is extremely hard to incorporate in analytic models. Only in recent years, thanks to major breakthroughs on analytical and numerical grounds, has our understanding of the Fermi process in
relativistic shocks \RecentA{significantly advanced}. This is the subject of the present review.

\RecentA{Relativistic shocks pose some unique challenges with respect to their non-relativistic counterparts.
For example, the distribution of accelerated particles can no longer be approximated as isotropic if the shock is relativistic.
In a relativistic shock, the electric and magnetic fields significantly mix as one switches between upstream and downstream frames of reference.
And unlike non-relativistic shocks, where some aspects of the theory can be tested by direct spacecraft measurements, relativistic shocks are only constrained by remote observations.
For recent reviews of relativistic shocks, see \cite{BykovTreumann11,2012SSRv..173..309B}.
}

This chapter is organized as follows.
First, we review recent analytical advances on the theory of particle acceleration in
relativistic shocks, arguing that the accelerated particle spectrum and its power-law slope in the ultra-relativistic limit, $s_\gamma\equiv -d\log N/d\log \gamma\simeq2.2$ (where $\gamma$ is the particle Lorentz factor), are fairly robust (Section \ref{particles}). Here, we assume {\it a priori} that some magnetic turbulence exists on both sides of the shock, such that the Fermi process can operate.
Next, we describe the plasma instabilities that are most relevant for generating this turbulence (Section \ref{waves}), stressing the parameter regime where the so-called Weibel (or ``filamentation'') instability -- which is often thought to mediate the Fermi process in weakly magnetized relativistic shocks -- can grow. Then, we summarize recent findings from particle-in-cell (PIC) simulations of
relativistic shocks, where the non-linear coupling between particles and magnetic waves can be captured from first principles (Section \ref{PIC}). Finally, we describe the astrophysical implications of these results for the acceleration of ultra high energy cosmic rays (UHECRs) and for the radiative signatures of PWNe and GRB afterglows (Section \ref{rad}; for a review of PWNe, see Kargaltsev et al. (2015) in the present volume; for a review of GRBs, see Racusin et al. (2015) in the present volume). We briefly conclude in Section \ref{conc}.

\section{Particle Acceleration in Relativistic Shocks}\label{particles}

Diffusive (Fermi) acceleration of charged particles in collisionless shocks is believed to be responsible for the production of non-thermal distributions of energetic particles in many astronomical systems \citep[][but see, \eg \cite{AronsTavani94} for a discussion of alternative shock acceleration processes]{blandford_eichler_87, MalkovDrury01}.
The Fermi acceleration process in shocks is still not understood from first principles: particle scattering in collisionless shocks is due to electromagnetic waves formed around the shock, but no present analytical formalism self-consistently calculates the generation of these waves, the scattering and acceleration of particles, and the backreaction of these particles on the waves and on the shock itself.

The theory of particle acceleration was first developed mainly by evolving the particle distribution under some Ansatz for the scattering mechanism (\eg diffusion in pitch angle), within the ``test particle'' approximation, where modifications of wave and shock properties due to the high energy particles are neglected.
This phenomenological approach proved successful in explaining the spectrum of relativistic particle distributions inferred from observations, although a more careful approach is needed to account for the energy fraction deposited in each particle species (electrons, positrons, protons, and possibly heavier ions), and to test the  Ansatz of the scattering prescription.

For \emph{non-relativistic} shocks, the linear theory of diffusive particle acceleration, first developed in 1977 \citep{Krymskii77, AxfordEtAl78, bell_78, blandford_ostriker_78}, yields a power-law distribution $d^3N/d^3\myp\propto \myp^{-\mysp}$ of particle momenta $\myp$, with a spectral index
\begin{equation} \mysp = s_\gamma+2 = 3\beta_u / (\beta_u-\beta_d) \, .  \label{eq:SIsoNR} \end{equation}
Here, $\beta$ is the fluid velocity normalized to the speed of light $c$ in the frame of the shock, which is assumed planar and infinite,
and subscripts $u$ ($d$) denote the upstream (downstream)
plasma. For strong shocks in an ideal gas of adiabatic index
$\Gamma=5/3$, this implies $\mysp=4$ (\ie $s_\gamma=2$; constant energy per
logarithmic energy interval, since $\myp^2d^3N/d^3\myp\propto \myp^{-2}$),
in agreement with observations.

The lack of a characteristic momentum scale, under the above assumptions, implies that the spectrum remains a power-law in the relativistic case, as verified numerically \citep{ostrowski_bednarz_98, achterberg_01}.
The particle drift downstream of the shock implies that more particles are moving downstream than upstream; this anisotropy is of order of $\beta_u$ when measured in the downstream frame \citep{keshet_waxman_05}.
Thus, while particle anisotropy is negligible for non-relativistic shocks, the distribution becomes highly anisotropic in the relativistic case, even when measured in the more isotropic downstream frame.
Consequently, one must simultaneously determine the spectrum and the angular distribution of the particles, which is the main difficulty underlying the analysis of test particle acceleration when the shock is relativistic.

Observations of GRB afterglows led to the conclusion that highly relativistic collisionless shocks  produce a power-law distribution of high energy particles with $\mysp=4.2\pm0.2$ \citep{Waxman97spectrum, FreedmanWaxman01, BergerEtAl03}. This triggered a numerical investigation of particle acceleration in such shocks, showing that $\mysp$ indeed approaches the value of $4.2$ for large shock Lorentz factors ($\gamma_{u}\equiv(1-\beta_u^2)^{-1/2}\gg1$), in agreement with GRB observations, provided that particle scattering is sufficiently isotropic.

The spectral index $\mysp$ was calculated under the test particle
approximation for a wide range of shock velocities, various equations
of state, and different scattering prescriptions.  This was achieved by
approximately matching numerical eigenfunctions of the transport equation between upstream and  downstream
\citep{KirkSchneider87, HeavensDrury88, kirk_00}, by Monte Carlo
simulations \citep{ostrowski_bednarz_98,
  achterberg_01,ellison_double_02,2003ApJ...589L..73L,niemiec_ostrowski_04,lemoine_revenu_06,EllisonEtAl13}, by expanding the distribution
parallel to the shock front \citep{keshet_waxman_05}, and by solving
for moments of the angular distribution \citep{Keshet06}.  

These studies have assumed rest frame diffusion in pitch angle or in the
angle between particle velocity and shock normal. These two
assumptions yield similar
spectra in the limit of ultra-relativistic shocks
\citep{ostrowski_bednarz_02}. {As discussed later in
  this review, one expects these assumptions to hold at relativistic
  shocks. However, some scenarios involve the conversion of the
  accelerated species into a neutral state and then back -- \eg proton to
  neutron and then back to proton via photo-hadronic interactions \citep{2003PhRvD..68d3003D} or
  electron to photon and then back to electron through Compton and pair
  production interactions \citep{stern_08} -- in which case the particle may have time
  to suffer a large angle deflection upstream of the shock, leading to
  large energy gains and generically hard
  spectra \citep{ostrowski_bednarz_98,MeliQuenby03, BlasiVietri05}.}

\begin{figure}[h]
\begin{center}
\includegraphics[width=0.75\textwidth]{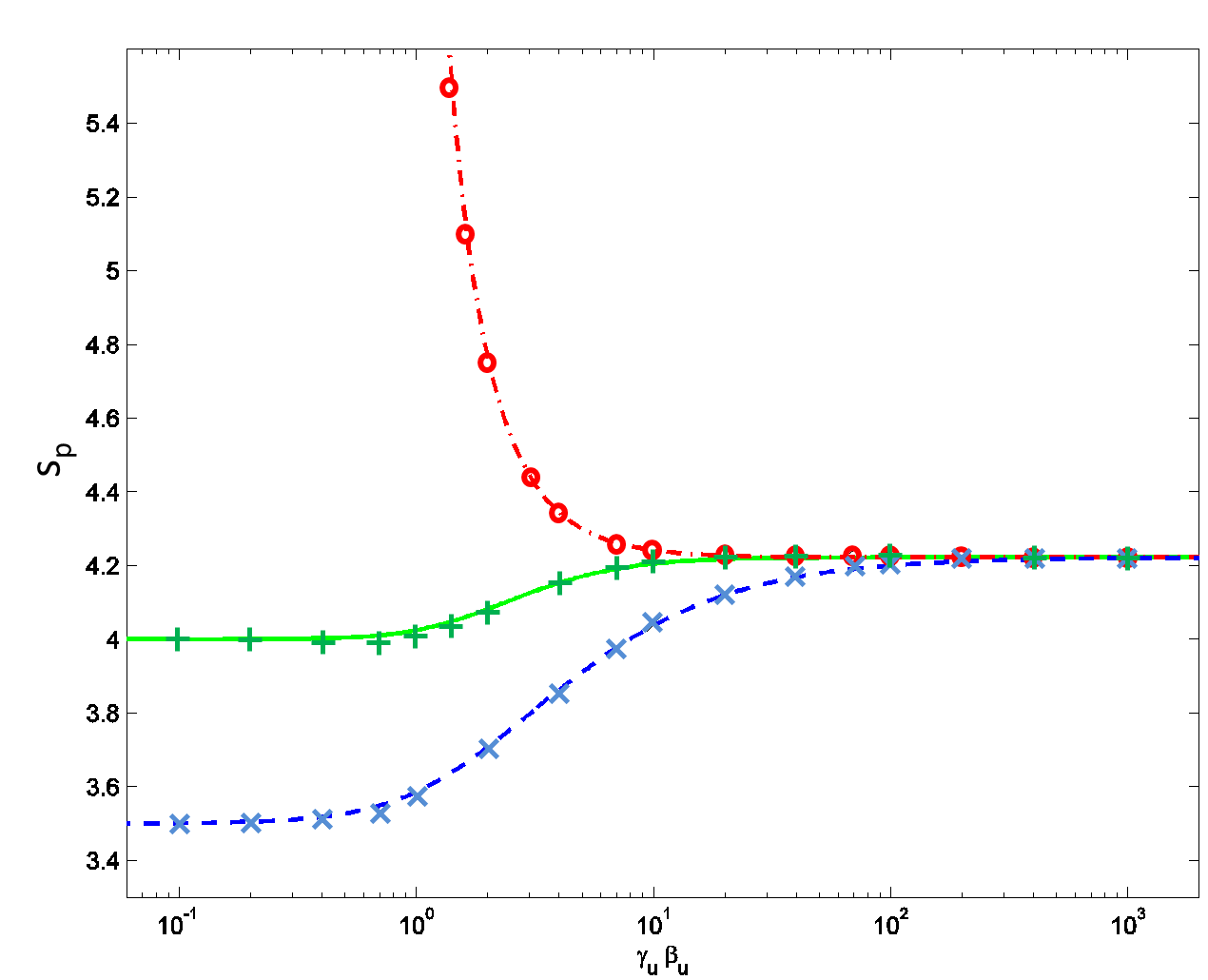}
\caption{\label{fig:SIso}
Spectral index according to Eq.~(\ref{eq:SIso}) \citep[][curves]{keshet_waxman_05} and to a numerical eigenfunction method \citep[][symbols]{kirk_00}, as a function of $\gamma_u \beta_u$, for three different types of shocks \citep{KirkDuffy99}: a strong shock with the J\"{u}ttner/Synge equation of state (solid curve and crosses), a strong shock with fixed adiabatic index $\Gamma=4/3$ (dashed curve and x-marks), and for a relativistic gas where $\beta_u \beta_d=1/3$ (dash-dotted curve and circles).
}
\label{fig:sh}
\end{center}
\end{figure}

For isotropic, small-angle scattering in the fluid frame, expanding the particle distribution about the shock grazing angle \citep{keshet_waxman_05} leads to a generalization of the non-relativistic Eq.~(\ref{eq:SIsoNR}) that reads
\begin{equation} \mysp = (3\beta_u - 2\beta_u \beta_d^2 + \beta_d^3) / (\beta_u - \beta_d) \, ,  \label{eq:SIso} \end{equation}
in agreement with numerical studies \citep{kirk_00, achterberg_01} over the entire range of $\beta_u$ and $\beta_d$.
In particular, in the ultra-relativistic shock limit, the spectral index becomes
\begin{equation} \mysp(\beta_u\rightarrow 1, \beta_d \rightarrow 1/3) = 38/9 = 4.222\ldots  \end{equation}
The spectrum is shown in \fig{SIso} for different equations of state, as a function of the shock four-velocity $\gamma_u\beta_u$.

The above analyses assumed that the waves scattering the particles move, on average, with the bulk fluid velocity.
More accurately, one should replace $\beta$ by the mean velocity of the waves that are scattering the particles.
In the shock precursor (see \S\ref{subsec:precursor}), the scattering waves are expected to be slower than the incoming flow, leading to a softer spectrum (smaller $\beta_u$ in Eq.~\ref{eq:SIso}).

Small-angle scattering can be parameterized by the angular diffusion function $\myD\equiv \langle(\Delta\theta)^2/\Delta t\rangle$, where $\theta$ is the angle of the particle velocity, taken here with respect to the shock normal, and angular brackets denote an ensemble average.
The function $\myD=\myD(\theta,\myp,z)$ should be specified on both sides of the shock, and in general depends on $\theta$, on the particle momentum $\myp$, and on its distance $z$ from the shock front.

For scattering off waves with a small coherence length $\lambda\ll
r_L$, where $r_L=(\myp c/eB)$ is the Larmor radius, roughly
$(r_L/\lambda)^2$ uncorrelated scattering events are needed in order
to produce an appreciable deflection, so $\myD\sim r_L^2 c /\lambda
\propto \myp^2$ \citep{achterberg_01,2009MNRAS.393..587P}.
Here, $B$ is the magnetic field, and $e$ is the electron's charge. 
Simulations \citep{sironi_13} confirm this scaling at early times; some implications are discussed
in Section \ref{PIC}.
The precise dependence of $\myD$ upon $z$ is not well known.
It is thought that $\myD$ slowly and monotonically declines away from the shock, as the energy in self-generated  fields decreases. However, the extents of the upstream precursor and downstream magnetized region are not well constrained observationally, and  in general are numerically unaccessible  in the foreseeable future.

For an evolved magnetic configuration, it is natural to assume that the diffusion function is approximately separable in the form $\myD=D(\theta)D_2(\myp,z)$.
Here, $D_2$ \citep[which may be approximately separable as well, but see][]{2007ApJ...655..375K} can be eliminated from the transport equation by rescaling $z$, such that the spectrum depends only on the angular part $D(\theta)$.

The spectrum is typically more sensitive to the downstream diffusion function $D_d$ than it is to the upstream $D_u$.
In general, an enhanced $D_d$ along (opposite to) the flow yields a softer (harder) spectrum; the trend is roughly reversed for $D_u$ \citep{Keshet06}.
Thus, the spectrum may deviate significantly from that of isotropic diffusion, in particular in the ultra-relativistic limit \citep{kirk_00, Keshet06}.
However, the spectral slope $s$ is not sensitive to localized changes in $D$ at angles perpendicular to the flow \citep{Keshet06}.
For roughly forward-backward symmetric scattering in the downstream frame, as suggested by PIC simulations, $s$ is approximately given by its isotropic diffusion value in Eq.~\ref{eq:SIso} (Keshet et al., in preparation).

Particle acceleration is thought to be efficient, at least in weakly magnetized or quasi-parallel shocks, as discussed below.
Thus, the relativistic particles are expected not only to generate waves, but also to slow down and heat the bulk plasma \citep{blandford_eichler_87}.
As particles with higher energies are expected to diffuse farther upstream and slow the plasma, lower-energy particles are effectively accelerated by a slower upstream.
Consequently, if the scattering waves are assumed to move with the bulk plasma, the  spectrum would no longer be a power-law.
However, this effect may be significant only for mildly relativistic shocks, with Lorentz factors below $\gamma_u\sim 3$ \citep{ellison_double_02, EllisonEtAl13}.

To understand the energy, composition, and additional features of the accelerated particles, such as the acceleration time and energy cutoffs, one must not only analyze the scattering of these particles (for example, by deriving $\myD$), but also address the injection problem, namely the process by which a small fraction of particles becomes subject to significant acceleration.
Such effects were investigated using Monte-Carlo techniques \citep{ostrowski_bednarz_98, EllisonEtAl13}, in the so-called ``thermal leakage'' model, where fast particles belonging to the downstream Maxwellian are assumed to be able to cross the shock into the upstream. More self-consistent results on particle injection based on PIC simulations are presented in Section \ref{PIC}. To uncover the physics behind the injection and acceleration processes, we next review the generation of electromagnetic waves in relativistic shocks.

\section{Plasma Instabilities in Relativistic Shocks}\label{waves}
\subsection{The Shock Precursor}\label{subsec:precursor}
\begin{figure}[!htb]
\begin{center}
\includegraphics[width=0.9\textwidth]{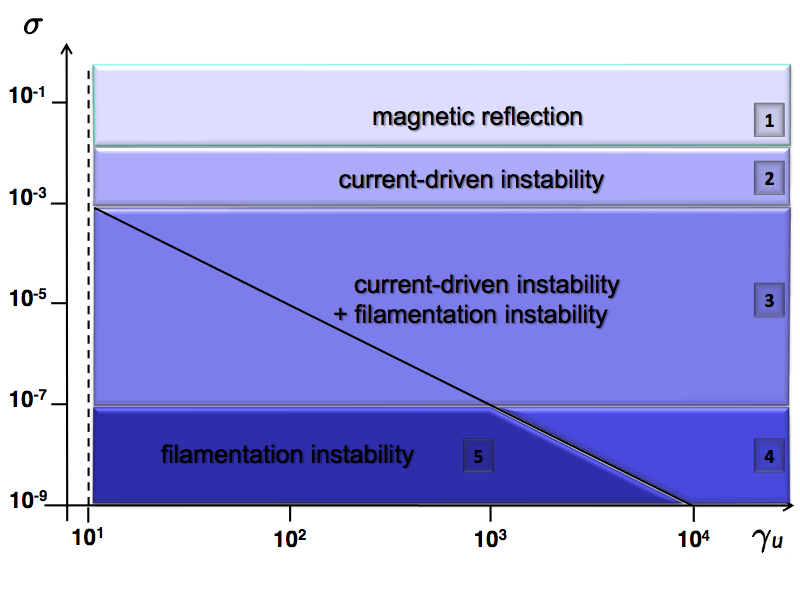}
\caption{\footnotesize{Phase diagram of relativistic collisionless shocks in the plane $(\gamma_u,\sigma)$; this figure assumes $\gamma_u>10$ and $\xi_{\rm cr}=0.1$, where the parameter $\xi_{\rm cr}\,=\,e_{\rm
    cr}/\left[\gamma_{u}(\gamma_{u}-1)n'mc^2\right]$
  characterizes the energy density of supra-thermal particles ($e_{\rm
    cr}$) relative to the incoming energy flux, as measured in the
  shock rest frame. In region 1, the shock transition is initiated by magnetic reflection in the compressed background field, while in regions $2-5$, the magnetic barrier is associated to the growth of micro-instabilities, as indicated. The solid diagonal line indicates values of $\sigma$ and $\gamma_u$ above which the filamentation instability would not have time to grow, in the absence of deceleration resulting from the compensation of the perpendicular current of the supra-thermal particles gyrating in the background field. See Section~3.1 and ~\cite{2014EL....10655001L} for a detailed discussion.}}
\label{martin}
\end{center}
\end{figure}

The collisionless shock transition is associated with the build-up of
some electromagnetic barrier, which is able to slow down and nearly isotropize
the incoming unshocked plasma. In media of substantial
magnetization\footnote{The magnetization is defined as
  $\sigma\,=\,B^2/\left[4\pi\gamma_{u}(\gamma_{u}-1)n'mc^2\right]$ in terms of $B$, the large-scale background magnetic
  field measured in the shock front rest frame, and $n'$, the proper
  upstream particle density. The mass $m$ is $m_p$ for an
  electron-proton shock, and $m_e$ for an electron-positron shock,
  \ie it corresponds to the mass of the particles which carry the
  bulk of the inertia. For a perpendicular shock, in which the background magnetic field in the shock frame is perpendicular to the flow, the magnetization
  can also be written as $\sigma\,=\,(u_{\rm A}/c)^2$, with $u_{\rm
    A}$ the Alfv\'en four-velocity of the upstream plasma.},
$\sigma\,\gtrsim\,10^{-2}$, this barrier can result from the
compression of the background magnetic field (as a result of the Lorentz transformation to the frame of a relativistic shock, the most generic configuration is that of a quasi-perpendicular field), while at lower
magnetizations, it is understood to arise from the generation of
intense micro-turbulence in the shock ``precursor'', as explained
hereafter and illustrated in Fig.~\ref{martin}.

At high magnetization, the gyration of the ambient particles in the
background compressed magnetic field can trigger a synchrotron maser
instability, which sends precursor electromagnetic waves into the
upstream~\citep{langdon_88,hoshino_91,hoshino_92,gallant_92}. As
incoming electrons and positrons interact with these waves, they undergo
heating~\citep{hoshino_08,sironi_spitkovsky_11a}, but acceleration
seemingly remains inefficient (Section~\ref{sec:mag}).

At magnetizations $\sigma\,\lesssim\,10^{-2}$, the interpenetration of
the incoming background plasma and the supra-thermal particles, which
have been reflected on the shock front or which are undergoing Fermi
cycles around the shock, leads to anisotropic micro-instabilities over
an extended region in front of the shock, called the ``precursor''
here. These instabilities then build up a magnetic barrier, up to a
level\footnote{The parameter $\epsilon_B$ denotes the magnetization of
  the turbulence, $\epsilon_B\,=\,\delta B^2/\left[4\pi\gamma_{
      u}(\gamma_{u}-1)n'mc^2\right]$, where $\delta B$ is the fluctuating magnetic field.}
$\epsilon_B\,\sim\,10^{-2}-10^{-1}$, sufficient to deflect strongly
the incoming particles and thus mediate the shock transition. This
picture, first envisioned by \citet{1963JNuE....5...43M}, has been recently
demonstrated in {\it ab initio} PIC
simulations~\citep{spitkovsky_05,2007ApJ...668..974K,spitkovsky_08}.
The generation of micro-turbulence in the shock precursor is thus a
key ingredient in the formation of the shock and in the development
of the Fermi process, as anticipated
analytically~\citep{2006ApJ...645L.129L} and from Monte Carlo
simulations~\citep{2006ApJ...650.1020N}, and demonstrated by PIC
simulations~\citep{spitkovsky_08b,sironi_spitkovsky_09,sironi_spitkovsky_11a},
see hereafter.

As seen in the background plasma (upstream) rest frame, the
supra-thermal particles form a highly focused beam, with an opening
angle $\sim\,1/\gamma_{u}$ and a mean Lorentz factor
$\overline\gamma_{\vert u}\,\sim\,\gamma_{u}^2$. In contrast, boosting
back to the shock frame, this supra-thermal particle distribution is now open
over $\sim\pi/2$, with a mean Lorentz factor $\overline\gamma_{\vert\rm
  sh}\,\gtrsim\,\gamma_{u}$, 
while the incoming plasma is highly focused,
with a mean Lorentz factor $\gamma_{u}$. A host of micro-instabilities
can in principle develop in such anisotropic configurations, see the
general discussion by \citet{bret_09}. However, in the
deep relativistic regime, the restricted length scale of the
precursor imposes a strong selection of potential instabilities, since
a background plasma volume element remains subject to the growth of
instabilities only while it crosses the
precursor. In the shock rest frame, this time scale is
$t_{\times,B}\,\simeq\,\omega_{\rm c}^{-1}$ in the presence of a
quasi-perpendicular background field\footnote{$\omega_{\rm
    c}\,\equiv\,e B_{\vert u}/m c$ represents the upstream frame
  cyclotron frequency (and $B_{\vert u}$ is the magnetic field in the
  upstream frame) while $\omega_{\rm p}\,\equiv\,\left(4\pi
  n'e^2/m\right)^{1/2}$ denotes the plasma frequency.} (a common field
geometry in relativistic flows), or $t_{\times,\delta
  B}\,\simeq\,\gamma_u\epsilon_B^{-1}\omega_{\rm
  p}^{-1}\left(\omega_{\rm p}\lambda_{\delta B}/c\right)^{-1}$ if the
scattering is dominated by short scale turbulence of magnetization
$\epsilon_B$ and coherence length $\lambda_{\delta B}$ (assuming that
the waves are purely magnetic in the rest frame of the background
plasma), see \eg \citet{2006ApJ...651..979M}, \citet{pelletier_10} and
\citet{plotnikov_12}. This small length scale implies that only the
fastest modes can grow, which limits the discussion to a few salient
instabilities.

Before proceeding further, one should stress that the above estimates
for $t_{\times}$ do not account for the influence of particles
accelerated to higher energies, which can propagate farther into the
upstream plasma and thus seed instabilities with smaller growth rate
and on larger spatial scales. While such particles do not carry the
bulk of the energy if the spectral index $s_\gamma>2$, it is anticipated that
they should nevertheless influence the structure of the precursor, see
in particular \citet{2006ApJ...651..979M},
\citet{2007ApJ...655..375K}, \citet{2009ApJ...696.2269M},
\citet{2009MNRAS.393..587P} and \citet{2014MNRAS.439.2050R} for
general analytical discussions, as well as \citet{keshet_09} for an
explicit numerical demonstration of their potential influence.
Similarly, the above estimates do not make a distinction between
electron-positron and electron-ion shocks; in particular, it is
understood that $\omega_{\rm c}$ and $\omega_{\rm p}$ refer to the
species which carries the bulk of the energy (\ie ions for electron-ion
shocks). PIC simulations have demonstrated that in electron-ion
shocks, electrons are heated in the precursor to nearly equipartition
with the ions, meaning that in the shock transition their relativistic
inertia becomes comparable to that of
ions~\citep[\eg][]{sironi_13}; hence one does not expect a
strong difference between the physics of electron-positron and
electron-ion shocks from the point of view of micro-instabilities, and
unless otherwise noted, this difference will be omitted in the
following. The microphysics of electron heating in the precursor
nevertheless remains an important open question, see
\citet{gedalin_08}, \citet{gedalin_12}, \citet{plotnikov_12} and
\citet{2015arXiv150105466K} for recent discussions of this issue;
indeed, the average Lorentz factor of electrons at the shock
transition directly impacts the peak frequency of the synchrotron
radiation of relativistic blast waves.

In the context of relativistic weakly magnetized shocks, the most
celebrated instability is the Weibel-like filamentation mode, which
develops through a charge separation in the background plasma,
triggered by magnetic fluctuations which segregate particles of
opposite charges into current filaments of alternate
polarity~\citep[\eg][]{gruzinov_waxman_99,medvedev_loeb_99,2004A&A...428..365W,2006ApJ...647.1250L,2007A&A...475....1A,2007A&A...475...19A,pelletier_10,2010PhRvE..81c6402B,2011ApJ...736..157R,pelletier_11,2011ApJ...738...93N,2012ApJ...744..182S}. The
current carried by the particles then positively feeds the magnetic
fluctuations, leading to fast growth, even in the absence of a net
large-scale magnetic field.  In the rest frame of the background
plasma, this instability grows in the linear regime as fast
as\footnote{The maximal growth rate of the Weibel instability is
  related to the plasma frequency of the beam of supra-thermal
  particles, $\omega_{\rm pb}$, though $\Im\omega \,\simeq\,\omega_{\rm
    pb}$, with $\omega_{\rm pb}\,\simeq\,\xi_{\rm cr}^{1/2}\omega_{\rm
    p}$.} $\Im\omega\,\simeq\,\xi_{\rm cr}^{1/2}\omega_{\rm p}$, with
maximum growth on scales of the order of $c/\omega_{\rm p}$; in the
filament rest frame, this instability is mostly of magnetic nature,
\ie $\Re\omega\,\sim\,0$.  Several branches of this instability have
been discussed in the literature, in particular the ``oblique mode'',
which involves a resonance with electrostatic modes. Even though this
latter mode grows slightly faster than the fully transverse
filamentation mode, it suffers from Landau damping once the electrons
are heated to relativistic temperatures, while the transverse
filamentation mode appears relatively insensitive to temperature
effects. Thus, at a first order approximation, the transverse filamentation mode
indeed appears to dominate the precursor at very low
magnetizations.  Its non-linear evolution, however, remains an open
question; analytical estimates suggest that it should saturate at
values $\epsilon_B\,\ll\,10^{-2}$ via trapping of the
particles~\citep{2004A&A...428..365W,2006ApJ...647.1250L,2007A&A...475....1A,2007A&A...475...19A},
while PIC simulations see a continuous growth of magnetic energy
density even when the non-linear filamentary structures have been
formed~\citep[\eg][]{keshet_09,sironi_13}. Whether
additional instabilities such as a kinking of the filaments contribute
in the non-linear phase thus remains debated, see for
instance~\citet{2006ApJ...641..978M}.

At moderate magnetization levels, another fast instability can be
triggered by the perpendicular current (transverse to both the
magnetic field and the shock normal) seeded in the precursor by the
supra-thermal particles during their gyration around the background
field~\citep{2014EL....10655001L,2014MNRAS.440.1365L}. The
compensation of this current by the background plasma on its entry into
the precursor leads to a deceleration of the flow, which modifies
somewhat the effective timescale available for the growth of plasma
instabilities, and destabilizes the modes of the background
plasma. The growth rate for this instability can be as large as
$\Im\omega\,\sim\,\omega_{\rm p}$, indicating that it can
compete with the Weibel filamentation mode at moderate
magnetizations. If the supra-thermal particle beam carries a net
charge (in the shock rest frame), or a net transverse current, other
similar instabilities are to be
expected~\citep[\eg][]{2009MNRAS.393..587P,2013MNRAS.433..940C,2014MNRAS.439.2050R}.
The phase space study of \citet{2014EL....10655001L} concludes that
the filamentation mode likely dominates at magnetization levels
$\sigma\,\lesssim\,10^{-7}$, while this perpendicular current-driven
instability dominates at
$10^{-3}\,\lesssim\,\sigma\,\lesssim\,10^{-2}$; in between, both
instabilities combine to form a complex precursor
structure. Interestingly, 
these results do not seem to depend on the
shock Lorentz factor, in good agreement with PIC
simulations~\citep{sironi_spitkovsky_09,sironi_spitkovsky_11a,sironi_13}.

Finally, one should mention the particular case of quasi-parallel
(subluminal) configurations: there, a fraction of the particles can in
principle escape to infinity along the magnetic field and seed other,
larger scale, instabilities. One prime candidate is the relativistic
generalization of the Bell streaming
instability~\citep[\eg][]{2006ApJ...651..979M,reville_06}, which is
triggered by a net longitudinal current of supra-thermal particles;
this instability has indeed been observed in PIC
simulations~\citep{sironi_spitkovsky_11a}. Of course, such a parallel
configuration remains a special case in the deep relativistic
regime. In mildly relativistic shock waves, with $\gamma_{
  u}\beta_{u}\,\sim\,1$, locally parallel configurations become
more frequent, hence one could expect such instabilities to play a key
role in seeding large scale turbulence.

\subsection{Downstream Magnetized Turbulence}  \label{sec:PIC_mag}
How the magnetized turbulence evolves downstream of the shock is an
important question, with direct connections to
observations.  The previous discussion suggests that the coherence
length of the fields generated in Weibel-like instabilities should be
comparable to the plasma skin-depth, $c/\omega_{\rm p}$. However, magnetic
power on such scales is expected to decay rapidly through
collisionless phase mixing~\citep{Gruzinov01}, while modeling of
GRB afterglow observations rather indicates that magnetic fields persist over scales $\sim\,10^7-10^9\,c/\omega_{\rm
  p}$ downstream~\citep{gruzinov_waxman_99}.

In a relativistic plasma, small-scale turbulence is dissipated at a
damping rate\footnote{The shock crossing conditions imply that the
  relativistic plasma frequency of the shocked downstream plasma is
  roughly the same as the plasma frequency of the upstream plasma; no
  distinction will be made here between these quantities.}
$\Im\omega\,\simeq\,-k^3c^3/\omega_{\rm
  p}^2$~\citep{chang_08,2015JPlPh..8145101L} as a function of the wavenumber
$k$, indicating that small scales are erased early on.
Larger modes can survive longer; 
power on scales exceeding the Larmor radius of the bulk plasma decays on long, $\Im\omega\,\propto k^2$ MHD scales \citep{keshet_09}. 
It is not clear at present whether the
small-scale turbulence manages to evolve to larger scales through
inverse cascade
effects~\citep[\eg][]{MedvedevEtAl05,2007ApJ...655..375K,2014ApJ...794L..26Z},
whether it is dissipated but at a rate which allows to match the
observations~\citep{lemoine_12,2013MNRAS.435.3009L}, or whether a
large-scale field is seeded in the downstream plasma by some external
instabilities~\citep[\eg][]{sironi_goodman_07,couch_08,2009ApJ...705L.213L}.

\begin{figure}[h]
\begin{center}
\includegraphics[width=0.75\textwidth]{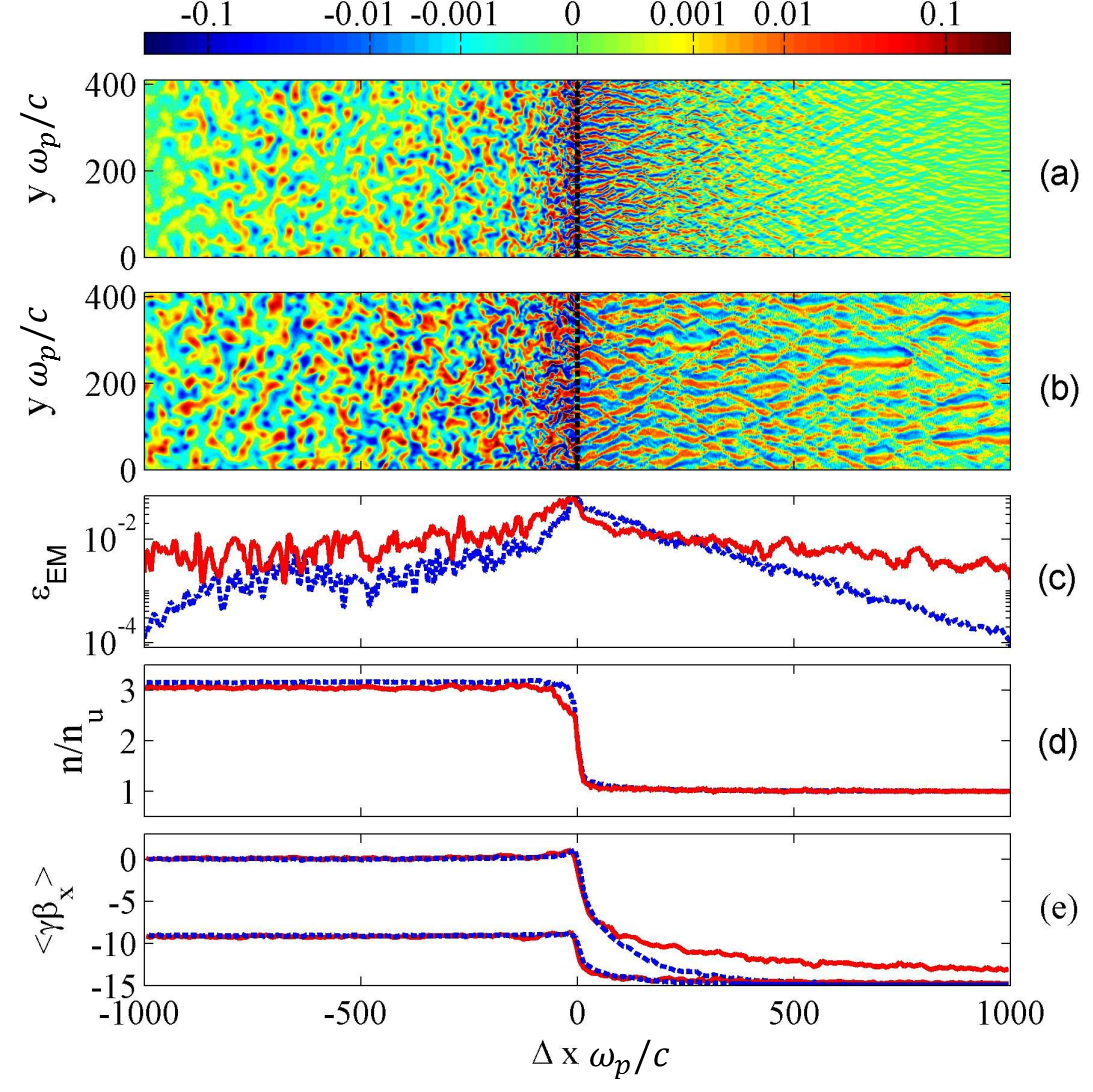}
\caption{ \footnotesize{Pair plasma evolution within $1000\,c/\omega_{\rm p}$ of the shock. {The simulation is performed in the downstream frame, and the upstream flow moves with a Lorentz factor $\gamma_r=15$ (so, $\gamma_r$ is the relative Lorentz factor between the upstream and downstream regions).}
  The normalized transverse magnetic field $\mbox{sign}(B)\,\epsilon_B$
  (color scale stretched in proportion to $\epsilon_B^{1/4}$ to
  highlight weak features) is shown at (a) early
  ($t_1=2250\,\omega_{\rm p}^{-1}$), and (b) late ($t_2=11925\,\omega_{\rm p}^{-1}$)
  times.  Here $\Delta x\equiv x-x_{\rm sh}$ is the distance from the
  shock, with $x_{\rm sh}$ (dashed vertical line) defined as the location of median density between far
  upstream and far downstream.  Also shown are the transverse averages
  (at $t_1$, dashed blue, and $t_2$, solid red) of (c) the electromagnetic
  energy $\epsilon_{EM} \equiv [(B^2+E^2)/8\pi]/[(\gamma_r-1)\gamma_rn'mc^2]$
  (with $E$ the electric field amplitude in the downstream frame, included in the definition of $\epsilon_{EM}$ because in the
  simulation frame the induced electric field in the upstream medium is $E\sim B$) normalized to the
  upstream kinetic energy, (d) density normalized to the far upstream density $n_u=\gamma_rn'$,
  and (e) particle momentum $\gamma\beta$ (with $\beta$ the velocity
  in $c$ units) in the x-direction averaged over all particles (higher
  $\ave{\gamma \beta_x}$) and over downstream-headed particles only.}}
\label{fig:PICMag}
\end{center}
\end{figure}

PIC simulations have quantified the generation of upstream current
filaments by pinching instabilities \citep[\eg][]{silva_03,
  frederiksen_04, JaroschekEtAl05, spitkovsky_05, spitkovsky_08,
  chang_08}, and resolved the formation of shocks in two- and three-dimensional (2D and 3D) pair
plasma \citep{spitkovsky_05, 2007ApJ...668..974K, chang_08, keshet_09,sironi_spitkovsky_09,haugbolle_10,sironi_spitkovsky_11a,sironi_13}
and ion-electron plasma \citep{spitkovsky_08,martins_09,sironi_13}. These simulations revealed a rapid
decay of the magnetic field downstream at early times
\citep{Gruzinov01, chang_08}.  Yet, a slow evolution of the plasma
configuration takes place on $>10^3/\omega_{\rm p}$ timescales, involving a
gradual increase in the scale of the magnetic structures, and
consequently their slower dissipation downstream \citep{keshet_09}.

This long-term evolution is driven entirely by the high-energy
particles accelerated in the shock; it is seen both upstream (\eg in the
precursor) and downstream, both of which become magnetized at
increasingly large distances from the shock, and with an increasingly
flat magnetic power-spectrum downstream \citep{keshet_09}. A flatter magnetic power
spectrum at the shock implies a larger fraction of the magnetic energy
stored in long-wavelength modes, which may survive farther from the shock.
Indeed, the index of a power-law spectrum of magnetic fluctuations
directly controls how fast the magnetic energy density, integrated over
wavenumbers, decays behind the
shock~\citep{chang_08,2015JPlPh..8145101L}; 
the scale-free limit corresponds to a flat magnetic power spectrum \citep{2007ApJ...655..375K}.

Properly capturing the backreaction of high energy particles requires
large simulation boxes and large particle numbers, to guarantee that
the largest scale fields and the highest energy particles are included.  The
largest available simulations at the present, with length $L$ and time $T$
scales of $(L\omega_{\rm p}/c)^2\, (T\omega_{\rm p}) \lesssim 10^{11}$, show no sign
of convergence at $T\gtrsim 10^4c/\omega_{\rm p}$ \citep{keshet_09,sironi_13}. This is illustrated in \fig{PICMag} for a pair-plasma shock
in 2D.

For magnetized shocks, the situation is different, as we describe below \citep{sironi_13}. At strong magnetizations, and for the quasi-perpendicular field geometry most relevant for relativistic flows, particle acceleration is suppressed, and the shock quickly reaches a steady state. At low (but nonzero) quasi-perpendicular magnetization, the shock evolves at early times similarly to the case of unmagnetized shocks (\ie $\sigma=0$). Particle acceleration proceeds to higher and higher energies, and modes of longer and longer wavelength appear. However, the maximum particle energy stops evolving once it reaches a threshold $\gamma_{sat}\propto \sigma^{-1/4}$  \citep{sironi_13}, and at that point the overall shock structure approaches a steady state.\citep{sironi_13}.\footnote{This conclusion regarding the saturation of the maximum particle Lorentz factor at $\gamma_{sat}$ has been tested in electron-positron shocks having $\sigma=\ex{4}-\ex{3}$ by \citet{sironi_13}, with the largest PIC study available to date. We caution that further nonlinear evolution, beyond the timespan covered by current PIC simulations, might be present in shocks with lower magnetization.}

\section{PIC Simulations of Relativistic Shocks}\label{PIC}
Only in the last few years, thanks to important advances in numerical algorithms and computer capabilities, plasma simulations have been able to tackle the problem of particle acceleration in relativistic shocks from first principles. In the following, we describe the major advances driven by large-scale PIC simulations in our understanding of particle acceleration in relativistic shocks. PIC codes can model astrophysical plasmas in the most fundamental way \citep{birdsall,buneman_93,spitkovsky_05}, as a collection of charged macro-particles that are moved by the Lorentz force. The currents deposited by the macro-particles on the computational grid are then used to solve for the electromagnetic fields via Maxwell's equations. The loop is closed self-consistently by extrapolating the fields to the macro-particle locations, where the Lorentz force is computed.

Full PIC simulations can capture, from first principles, the acceleration physics of both electrons and ions. However, such simulations must resolve the electron plasma skin depth $c/\ompe$, which is typically much smaller than astrophysical scales. Hence, most simulations can only cover limited time and length scales, and usually with low dimensionality (1D or 2D instead of 3D) and small ion-to-electron mass ratios (the ion skin depth $c/\omega_{\rm pi}$ is a factor of $\sqrt{m_i/m_e}$ larger than the electron skin depth $c/\ompe$). The results discussed below pertain to simulation durations of order $\sim10^3-10^4\,\ompe$ in electron-positron shocks and $\sim10^3\,\omega_{\rm pi}$ in electron-ion shocks (but with reduced mass ratios), 
so a careful extrapolation is needed to bridge these microscopic scales with the macroscopic scales of astrophysical interest. Yet, as we review below, PIC simulations provide invaluable insight into the physics of particle injection and acceleration in astrophysical sources.

The structure of relativistic shocks and the efficiency of particle acceleration depend on the conditions of the upstream flow, such as bulk velocity, magnetic field strength and field orientation. PIC simulations have shown that the shock physics and the efficiency of particle acceleration are insensitive to the shock Lorentz factor (modulo an overall shift in the energy scale), in the regime $\gamma_r\gg1$ of ultra-relativistic flows \citep[\eg][]{sironi_13}. Below, we only discuss results for shocks where the upstream Lorentz factor with respect to the downstream frame is $\gamma_r\gtrsim 5$, neglecting the trans- and non-relativistic regimes that are outside the scope of this review. We discuss the physics of both electron-positron shocks and electron-ion shocks (up to realistic mass ratios), neglecting the case of electron-positron-ion shocks presented by \eg \citet{hoshino_92,amato_arons_06,stockem_12}, which might be relevant for PWNe.
As found by \citet{sironi_spitkovsky_09,sironi_spitkovsky_11a,sironi_13}, for highly relativistic flows, the main parameter that controls the shock physics is the magnetization $\sigma$. Below, we distinguish between shocks propagating into strongly magnetized media ($\sigma\gtrsim \ex{3}$) and weakly magnetized or unmagnetized shocks ($\sigma\lesssim \ex{3}$).

\subsection{Particle Acceleration in Strongly Magnetized Shocks}\label{sec:mag}
\begin{figure}[!htb]
\begin{center}
\includegraphics[width=0.8\textwidth]{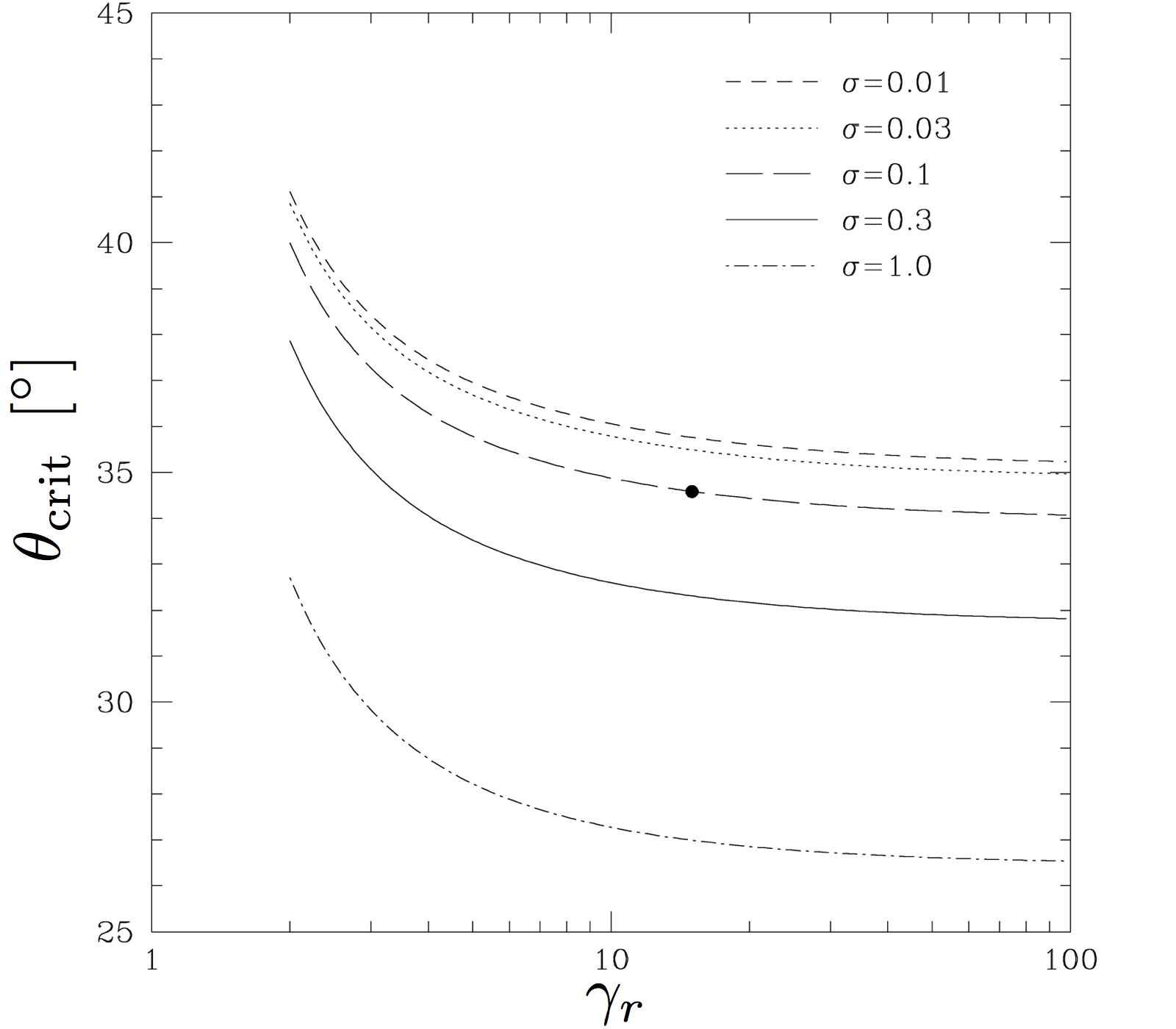}
\caption{\footnotesize{Critical obliquity angle $\theta_{\rm crit}$ (measured in the downstream frame) that separates subluminal and superluminal configurations \citep{sironi_spitkovsky_09}, as a function of the flow Lorentz factor $\gamma_r$ and the magnetization $\sigma$, as indicated in the label. The filled black circle indicates our reference case with $\gamma_r=15$ and $\sigma=0.1$.}}
\label{fig:super}
\end{center}
\end{figure}

For high magnetizations ($\sigma\gtrsim10^{-3}$ in electron-positron flows, or $\sigma\gtrsim3\times10^{-5}$ in electron-ion flows), the shock structure and acceleration properties depend critically on the inclination angle $\theta$ between the upstream field and the shock direction of propagation \citep{sironi_spitkovsky_09,sironi_spitkovsky_11a}. If the magnetic obliquity is larger than a critical angle $\theta_{\rm crit}$, charged particles would need to move along the field faster than the speed of light in order to outrun the shock (``superluminal'' configurations). In \fig{super}, we show how the critical angle $\theta_{\rm crit}$ (as measured in the downstream frame) depends on the flow velocity and magnetization. In the limit of $\sigma\ll1$ and $\gamma_r\gg1$, the critical obliquity approaches the value $\theta_{\rm crit}\simeq34^\circ$.

\begin{figure}[!tbp]
\begin{center}
\PNGfigure{\includegraphics[width=1\textwidth]{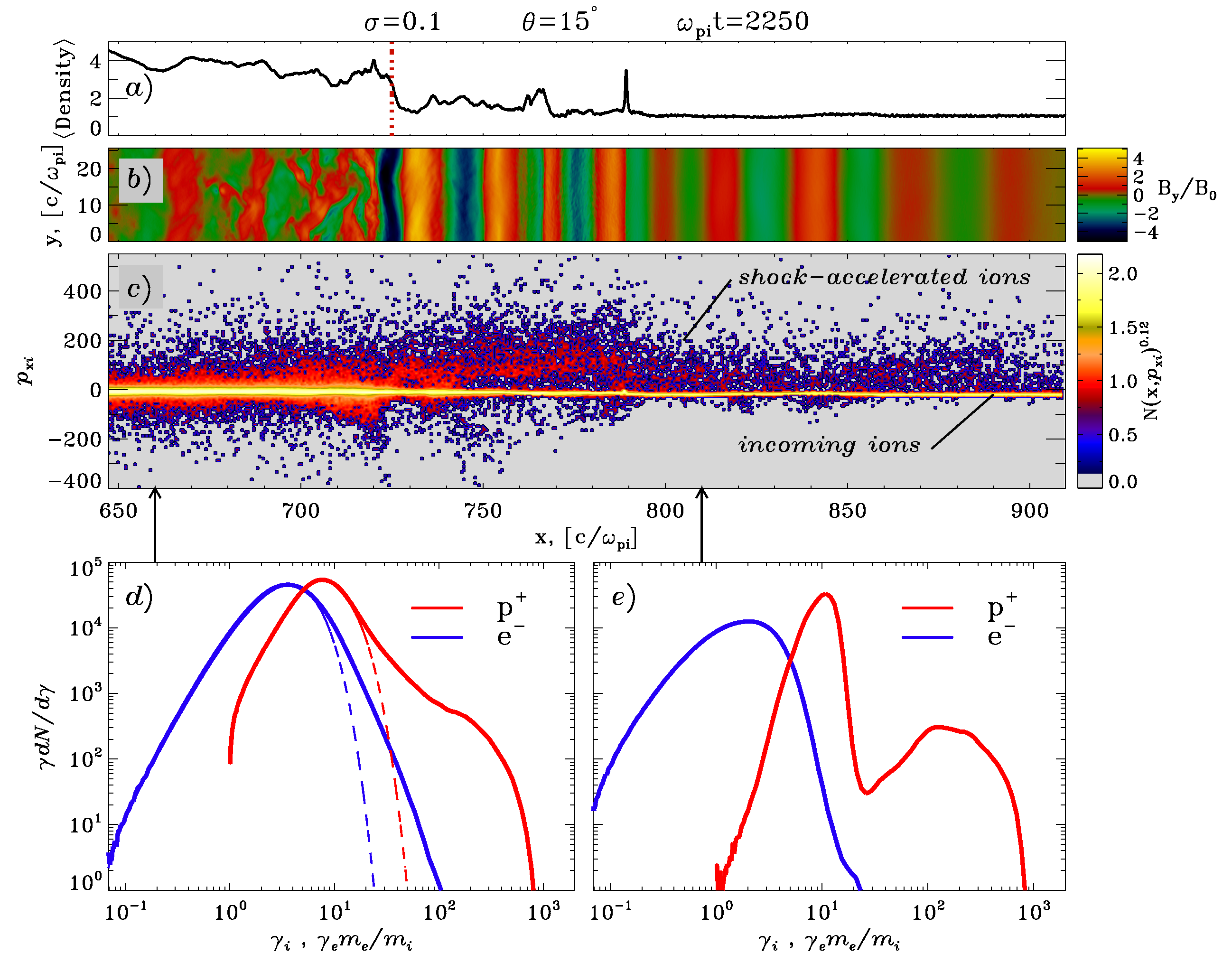}}
\caption{\footnotesize{Structure of an electron-ion subluminal shock with $\gamma_r=15$, $\sigma=0.1$ and $\theta=15^\circ$, from \citet{sironi_spitkovsky_11a}. The simulation is performed in the downstream frame. The shock front is located at $x\sim725\,c/\omega_{\rm pi}$ (vertical dotted red line in panel (a)), and it separates the upstream region (to its right) from the compressed downstream region (to its left). A stream of shock-accelerated ions propagates ahead of the shock (see the diffuse cloud in the momentum space $x-p_{xi}$ of panel (c) to the right of the shock, at $x\gtrsim725\,c/\omega_{\rm pi}$). Their interaction with the upstream flow (narrow beam to the right of the shock in panel (c)) generates magnetic turbulence ahead of the shock (see the transverse waves in panel (b), to the right of the shock). In turn, such waves govern the process of particle acceleration. In fact, the particle spectrum behind the shock (solid lines in panel (d); red for ions, blue for electrons) is not compatible with a simple thermal distribution (dashed lines), showing a clear non-thermal tail of high-energy particles, most notably for ions.} }
\label{fig:shock1}
\end{center}
\end{figure}

Only ``subluminal'' shocks ($\theta\lesssim\theta_{\rm crit}$) are efficient particle accelerators \citep{sironi_spitkovsky_09,sironi_spitkovsky_11a,sironi_13}, in agreement with the analytical findings of \citet{begelman_kirk_90}. As illustrated in Fig.~\ref{fig:shock1}, a stream of shock-accelerated particles propagates ahead of the shock (panel (c)), and their counter-streaming with the incoming flow generates magnetic turbulence in the upstream region (panel (b)). In turn, such waves govern the acceleration process, by providing the turbulence required for the Fermi mechanism. In the particular case of \fig{shock1} --- a relativistic shock with $\gamma_r=15$, $\sigma=0.1$ and $\theta=15^\circ$ propagating  into an electron-ion plasma --- the upstream turbulence is dominated by Bell-like modes \citep{reville_06,pelletier_10,pelletier_11}.
The downstream particle spectrum in subluminal shocks shows a pronounced non-thermal tail of shock-accelerated particles with a power-law index $2\lesssim s_\gamma\lesssim 3$ (panel (d)). The tail contains $\sim5\%$ of the particles and $\sim20\%$ of the flow energy at time $2250\,\omega_{\rm pi}^{-1}$; both values appear to be time-converged, within the timespan covered by our simulations.

In contrast, superluminal shocks ($\theta\gtrsim\theta_{\rm crit}$) show negligible particle acceleration \citep{gallant_92,hoshino_08,sironi_spitkovsky_09,sironi_spitkovsky_11a,sironi_13}. Here, due to the lack of significant self-generated turbulence, charged particles are forced to slide along the background field lines, whose orientation prohibits repeated crossings of the shock. This inhibits the Fermi process, and in fact the particle distribution behind superluminal shocks is purely thermal. The same conclusion holds for both electron-positron and electron-ion flows. In electron-ion shocks, the incoming electrons are heated up to the ion energy, due to powerful electromagnetic waves emitted by the shock into the upstream medium, as a result of the synchrotron maser instability (studied analytically by \citet{lyubarsky_06}, and with 1D PIC simulations by \eg\citet{langdon_88,gallant_92,hoshino_92,hoshino_08}). Yet, such heating is not powerful enough to permit an efficient injection of electrons into the Fermi acceleration process at superluminal electron-ion shocks.

If magnetized superluminal shocks are responsible for producing the radiating particles in astrophysical relativistic sources, the strong electron heating  observed in electron-ion shocks implies that the putative power-law tail in the electron spectrum should start from energies higher than the ion bulk kinetic energy. For models of GRBs and AGN jets that require a power-law distribution extending down to lower energies, the presence of such shocks would suggest that electron-positron pairs may be a major component of the flow.

\begin{figure}[!htb]
\begin{center}
\label{fig:tot}\includegraphics[width=1.3\textwidth,angle=0]{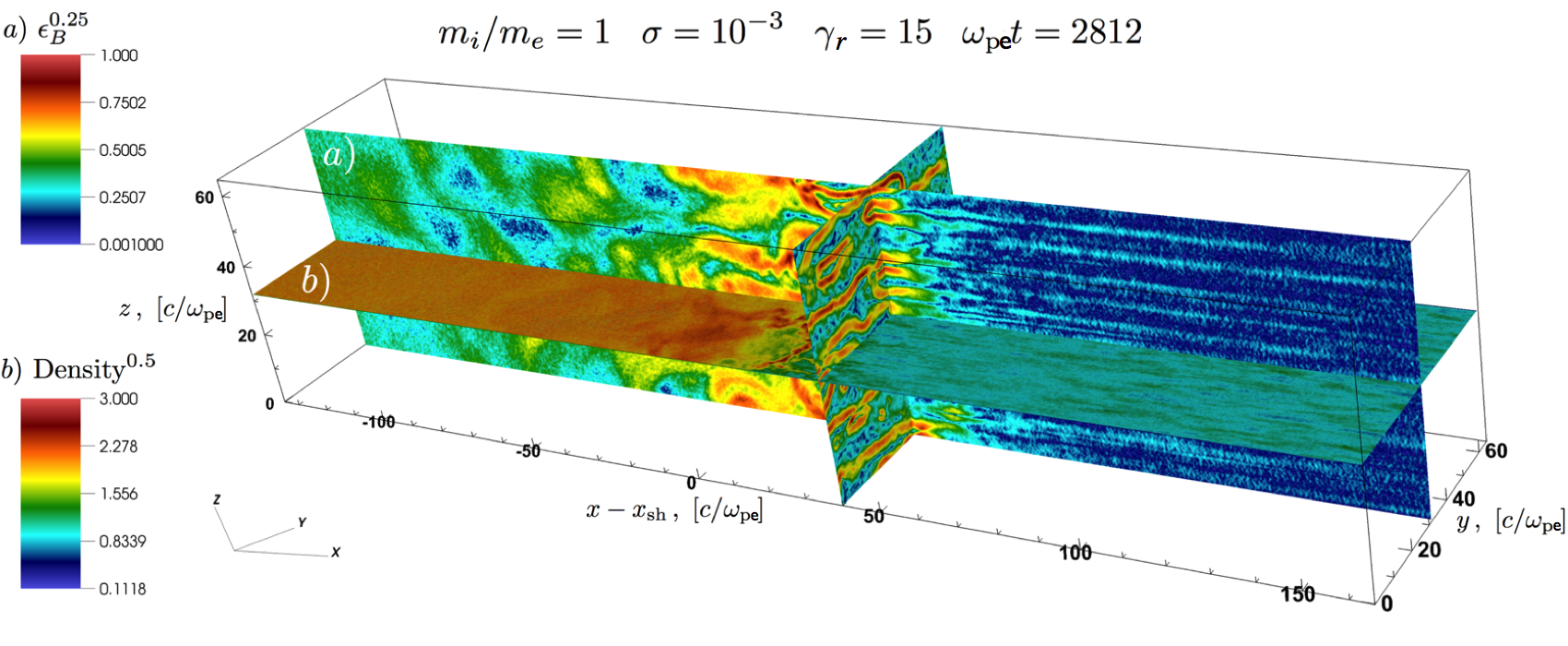}
\caption{\footnotesize{Shock structure from the 3D PIC simulation of a $\sigma=10^{-3}$ electron-positron shock with $\gamma_r=15$, from \cite{sironi_13}. The simulation is performed in the downstream frame and the shock propagates along $+\hat{x}$. We show the $xy$ slice of the particle number density (normalized to the upstream density), and the $xz$ and $yz$ slices of the magnetic energy fraction $\epsilon_B$. A stream of shock-accelerated particles propagates ahead of the shock, and their counter-streaming motion with respect to the incoming flow generates magnetic turbulence in the upstream via electromagnetic micro-instabilities. In turn, such waves provide the scattering required for particle acceleration.}}
\label{fig:shock}
\end{center}
\end{figure}

\subsection{Particle Acceleration in Weakly Magnetized and Unmagnetized Shocks}
Weakly magnetized shocks ($\sigma\lesssim10^{-3}$ in electron-positron flows, $\sigma\lesssim3\times10^{-5}$ in electron-ion flows) are governed by electromagnetic plasma instabilities (see \S\ref{subsec:precursor}), that generate magnetic fields stronger than the background field. Such shocks do accelerate particles self-consistently, regardless of the magnetic obliquity angle \citep[][]{spitkovsky_08,spitkovsky_08b,martins_09,haugbolle_10,sironi_13}. The stream of shock-accelerated particles propagates ahead of the shock, triggering the Weibel instability. The instability generates filamentary magnetic structures in the upstream region, as shown in Fig.~\ref{fig:shock}, which in turn scatter the particles back and forth across the shock, mediating Fermi acceleration.

\begin{figure}[!hbt]
\begin{center}
\includegraphics[width=0.75\textwidth]{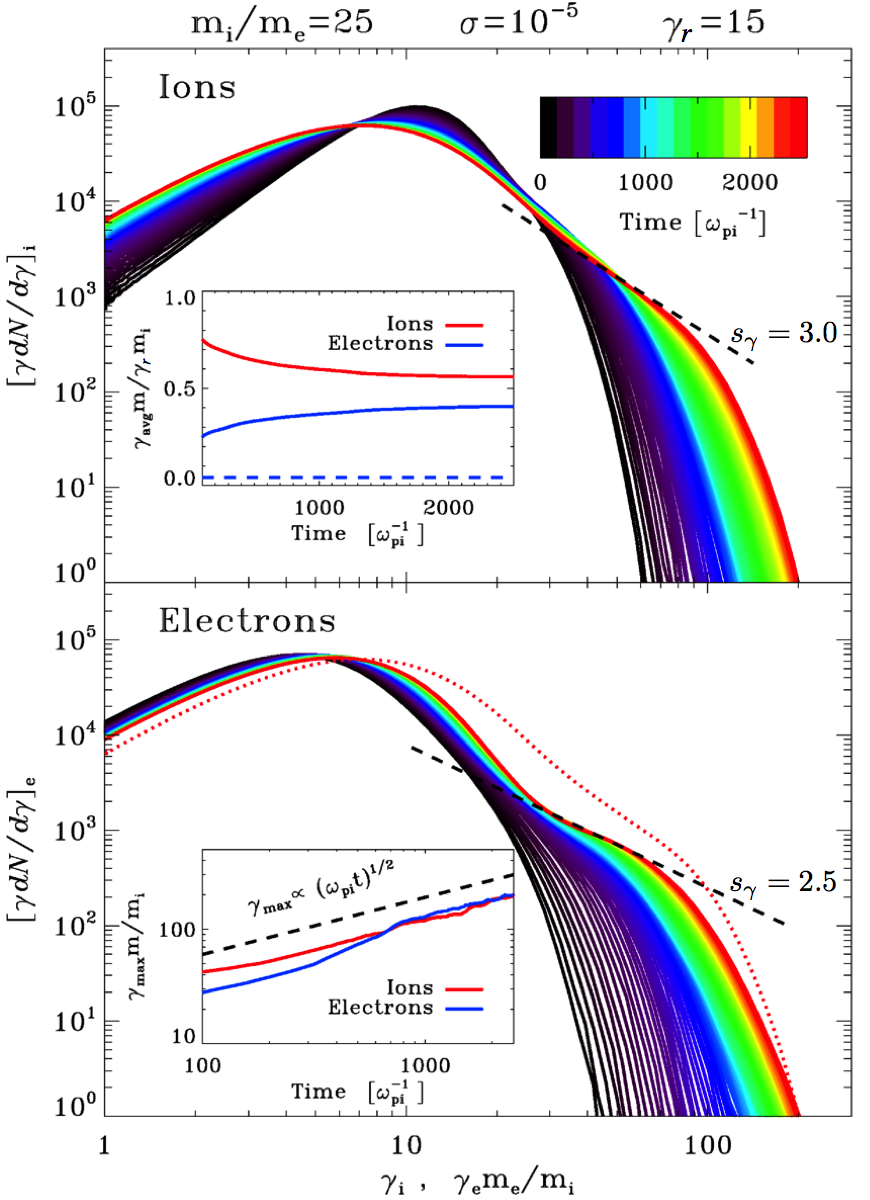}
\caption{\footnotesize{Temporal evolution of the downstream particle spectrum, from the 2D simulation of a $\gamma_r=15$ electron-ion ($m_i/m_e=25$) shock propagating into a flow with magnetization $\sigma=10^{-5}$, from \citet{sironi_13}. The evolution of the shock is followed from its birth (black curve) up to $\omega_{\rm pi}t=2500$ (red curve). In the top panel we show the ion spectrum and in the bottom panel the electron spectrum. The non-thermal tails approach at late times a power law with a slope $s_\gamma=3.0$ for ions and $s_\gamma=2.5$ for electrons (black dashed lines in the two panels). In the bottom panel, we overplot the ion spectrum at $\omega_{\rm pi}t=2500$ with a red dotted line, showing that ions and electrons are nearly in equipartition. Inset of the top panel: mean downstream ion (red) and electron (blue) energy, in units of the bulk energy of an upstream particle. The dashed blue line shows the electron energy at injection. Inset of the bottom panel: temporal evolution of the maximum Lorentz factor of ions (red) and electrons (blue), scaling as $\propto (\omega_{\rm pi} t)^{1/2}$ at late times (black dashed line).}}
\label{fig:accel1}
\end{center}
\end{figure}

The accelerated particles in weakly magnetized shocks populate in the downstream region a power-law tail $dN/d\gamma\propto \gamma^{-s_\gamma}$ with a slope $s_\gamma\sim2.5$, that contains $\sim3\%$ of the particles and $\sim10\%$ of the flow energy.\footnote{These values are nearly independent of the flow composition and magnetization, in the regime of weakly magnetized shocks. Also, they are measured at time $\sim 10^4\,\omega_{\rm pe}^{-1}$ in electron-positron shocks and at $\sim 10^3\,\omega_{\rm pi}^{-1}$ in electron-ion shocks, but they appear remarkably constant over time,  within the timespan covered by our simulations.} In electron-ion shocks, the acceleration process proceeds similarly for the two species, since the electrons enter the shock nearly in equipartition with the ions, as a result of strong pre-heating in the self-generated Weibel turbulence \citep{spitkovsky_08,martins_09,sironi_13}. In both electron-positron and electron-ion shocks, the maximum energy of the accelerated particles scales in time as $\gamma_{max}\propto t^{1/2}$ \citep{sironi_13}, as shown in Fig.~\ref{fig:accel1}. More precisely, the maximum particle Lorentz factor in the downstream frame scales as
\be\label{eq:gmax4a}
&\gamma_{max}&\simeq0.5\,\gamma_r\,(\ompt)^{1/2}\\
\gamma_{max,i}&\sim\frac{\gamma_{max,e} m_e}{m_i}&\simeq0.25\,\gamma_r\,(\omega_{\rm pi}t)^{1/2}\label{eq:gmax4b}
\ee
in electron-positron and in electron-ion shocks, respectively \citep{sironi_13}. This scaling is shallower than the so-called (and commonly assumed) Bohm limit $\gamma_{max}\propto t$, and it naturally results from the small-scale nature of the Weibel turbulence generated in the shock layer (see Fig.~\ref{fig:shock}).

\begin{figure}[!htb]
\begin{center}
\includegraphics[width=0.75\textwidth]{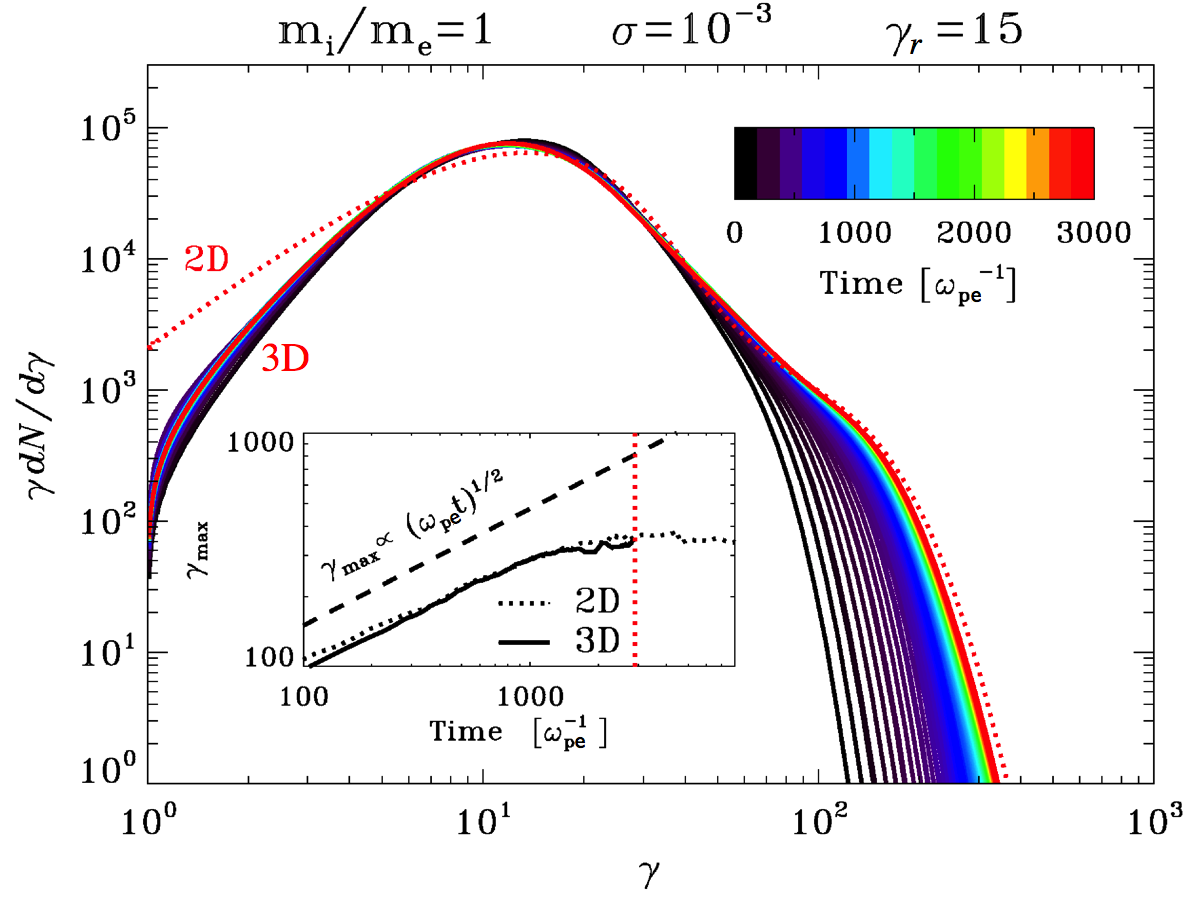}
\caption{\footnotesize{Time evolution of the downstream particle spectrum from the 3D PIC simulation of a $\sigma=10^{-3}$ electron-positron shock with $\gamma_r=15$, from \cite{sironi_13}. The evolution of the shock is followed from its birth (black curve) up to $\omega_{\rm pe}t=3000$ (red curve). We overplot the spectrum at $\omega_{\rm pe}t=3000$ from a 2D simulation with the same parameters (red dotted line), showing excellent agreement at high energies. The inset shows that the maximum particle Lorentz factor grows as $\gamma_{max}\propto t^{1/2}$, before saturating at $\gamma_{sat}\propto \sigma^{-1/4}$. The results are consistent between 2D (dotted) and 3D (solid).}}
\label{fig:accel2}
\end{center}
\end{figure}

The increase of the maximum particle energy over time proceeds up to a saturation Lorentz factor (once again, measured in the downstream frame) that is constrained by the magnetization $\sigma$ of the upstream flow
\be\label{eq:gsat4a}
&\gamma_{sat}&\simeq4\,\gamma_r\,\sigma^{-1/4}\\
\gamma_{sat,i}&\sim\frac{\gamma_{sat,e} m_e}{m_i}&\simeq2\,\gamma_r\,\sigma^{-1/4}\label{eq:gsat4b}
\ee
in electron-positron and electron-ion shocks, respectively. The saturation of the maximum particle energy is shown in Fig.~\ref{fig:accel2} for a shock with $\sigma=\ex{3}$. Further energization is prevented by the fact that the self-generated turbulence is confined within a region of thickness $L_{B,sat}\propto \sigma^{-1/2} $ around the shock \citep{sironi_13}.

\section{Astrophysical Implications}\label{rad}

\subsection{Acceleration of Ultra-High Energy Cosmic Rays}
Relativistic shock waves have long been considered as prime candidates
for the acceleration of cosmic rays to the highest energies observed,
$E\,\sim\,10^{20}\,$eV. Indeed, a naive extrapolation of the
acceleration time scale in the sub-relativistic regime ($t_{\rm
  acc}\,\sim\,t_{\rm scatt}/\beta_{u}^2$, with $t_{\rm scatt}$
the scattering timescale) suggests that relativistic shocks (\ie $\beta_u\sim 1$) accelerate
particles on shorter time scales than non-relativistic shocks (\ie $\beta_u\ll 1$), at a given $t_{\rm scatt}$. For given radiative loss and escape time
scales, this implies that relativistic shocks would be accelerating particles to much higher energies than non-relativistic shocks. However, the situation is more complex than it appears; in
particular, in relativistic shock waves, $t_{\rm scatt}$ may be much
larger than usually assumed.

As mentioned repeatedly in the previous paragraphs, particle
acceleration in the relativistic regime $\gamma_{u}\beta_{
  u}\,\gg\,1$ around a steady planar shock wave, operates only if
intense micro-turbulence has been excited in the shock precursor, as
demonstrated analytically~\citep{2006ApJ...645L.129L}, by Monte Carlo
simulations~\citep{2006ApJ...650.1020N} and by PIC
simulations~\citep{sironi_13}; consequences for the
acceleration of particles to ultra-high energies have been discussed
in several papers, \eg by  \citet{2008AIPC.1085...61P},
\citet{2009JCAP...11..009L}, \citet{2011AIPC.1367...70L},
\citet{2011ApJ...738L..21E}, \citet{2012SSRv..173..309B,sironi_13} or more
recently by \citet{2014MNRAS.439.2050R}.

Scattering in small-scale turbulence leads to a downstream residence
time $t_{\rm scatt}\,\sim\,r_{L}^2/(\lambda_{\delta B}c)$, with
$r_{L}$ the Larmor radius of the particle and $\lambda_{\delta B}$
the coherence length scale of the turbulence. This implies that the
(shock frame) acceleration timescale $t_{\rm acc}$ grows quadratically
with the energy, which fits well the result seen in PIC simulations
that the maximum energy grows as the square root of time. In other
words, as the particle energy grows, the acceleration timescale
departs more and more from the Bohm estimate, which is generally used
to compute the maximum energy. Comparing for instance the acceleration
timescale, which is at least equal to the above downstream residence time, with the dynamical timescale $r/\gamma_{u}$  in
the shock rest frame ($r$
denoting the radius of the shock front in the upstream rest frame), one finds a maximum energy $E_{\rm max}\,\lesssim \,
e\, \delta B\, r\left(\gamma_{u}\lambda_{\delta B}/r\right)^{1/2}$,
with $\delta B$ the strength of the turbulent field expressed in the
shock frame; the above maximal energy has been written in the upstream
(observer) frame. The factor in the brackets generally takes very
small values, because $\lambda_{\delta B}\,\sim\,c/\omega_{\rm p}$
while $r$ is a macroscopic length scale; this maximal energy is thus
far below the so-called Hillas estimate $e\,\delta B r$, which
corresponds to a Bohm estimate for $t_{\rm scatt}$.

Another way to phrase the problem is as follows (see
\citealt{2009JCAP...11..009L} for a discussion): assume that the
acceleration timescale is written $t_{\rm acc}\,=\,{\cal A}\,r_{L}/c$, and derive the maximum energy by comparing $t_{\rm acc}$ with
$t_{\rm dyn}\,=\,r/(\gamma\beta c)$ as above for a jet moving at
velocity $\beta$ towards the observer. Then one finds that acceleration of particles of
charge $Z$ to $10^{20}E_{20}\,$eV requires that the isotropic
equivalent magnetic luminosity of the object exceeds:
$L_B\,\gtrsim\,10^{45}\,\,Z^{-2}E_{20}^2{\cal A}^2\gamma^2\,$erg/s, a
very large number indeed, all the more so if ${\cal A}\,\gg\,1$. For
acceleration at ultra-relativistic shock waves,
${\cal A}$ is much larger than unity (while the Bohm estimate corresponds to ${\cal A}\,\sim\,1$), with typical values ${\cal A}\,\sim\,E/\left(\gamma_u m_p c^2\right)$.
In summary, particle acceleration at ultra-relativistic shock waves
does not appear fast enough to produce particles of ultra-high
energies. In particular, when the above arguments are applied to the
case of the external shock of a GRB, the maximal energy is
found to be of the order of
$10^{16}\,$eV~\citep{plotnikov_12,sironi_13,2014MNRAS.439.2050R}.

It is important however to note three caveats in the above arguments.
One is that as $\gamma_{u}\beta_{u}\,\rightarrow\,1$, \ie for mildly relativistic shock waves, the
nature of the turbulence remains unknown and one cannot exclude
that scattering would be closer to a Bohm estimate. Two facts support
such a speculation: (1) the precursor increases in size as
$\gamma_{u}$ diminishes, which suggests that MHD-scale
instabilities could arise and excite large scale turbulence; and (2),
the obliquity becomes less of a problem for mildly relativistic shock
waves, suggesting that large scale turbulence could possibly lead to
acceleration in this regime. A second caveat is the fact that PWNe are very efficient particle accelerators, even though one
would expect the opposite in the absence of reconnection or other
dissipative processes, due to the large magnetization of the flow
(Section~\ref{sect:pwn}). More precisely, synchrotron photons are
observed with energies as high as $100\,$MeV, which means that pairs
are accelerated up to the radiation-reaction limit, \ie with an
acceleration time scale close to the theoretical Bohm scaling. Such
empirical evidence suggests that ions could also be accelerated to
very high energies, if ions are indeed injected along with pairs in
the wind. In the Crab, such a maximal energy would be limited by the
confinement in the nebular turbulence to values of the order of
$10^{17}\,$eV (for $Z=1$); more powerful nebulae, associated with
young pulsars born with a few millisecond periods, could however confine (and
potentially accelerate) protons up to the highest energies
\citep{2014arXiv1409.0159L}.
Finally, as the nonlinear evolution of weakly magnetized or parallel shocks over long timescales is not yet understood, some of the above estimates, pertaining \eg to the diffusive properties and extent of the magnetic field, may be altered on macroscopic times.

\subsection{Radiative Signatures of Relativistic Blast Waves}
In line with the previous discussion, one can compute the maximal
energy for electrons and derive the maximal synchrotron photon
energy. Using an acceleration time scale $t_{\rm acc}\,\simeq\,r_{L}^2/(\lambda_{\delta B}c)$ and comparing to synchrotron losses in
the self-generated micro-turbulence, characterized by its
magnetization $\epsilon_B$, one derives a maximum synchrotron photon
energy of the order of a few GeV in the early phase of GRB
afterglows, \ie during the first hundreds of
seconds~\citep{kirk_reville_10,plotnikov_12,lemoine_12,2013ApJ...771L..33W,sironi_13}. Let
us stress that in the latter study, this estimate has been derived
from PIC simulations with a self-consistent measurement of the
acceleration time scale in the self-consistent magnetic field. The
synchrotron radiation of electrons accelerated at the external
ultra-relativistic shock of GRBs can thus produce the bulk
of the long-lasting $>100\,{\rm MeV}$ emission detected by the Fermi
satellite \citep[\eg][]{barniol_09,ackermann_10,depasquale_10,ghisellini_10}. The photons that have been observed with energies in excess
of $\gtrsim10\,$GeV probably result from inverse Compton
interactions~\citep{2013ApJ...771L..33W}. Interestingly, the recent
GRB130427A has revealed a long-lasting emission with a possible break
in the spectrum at an energy of a GeV, characteristic of a turn-over
between the synchrotron and the synchrotron self-Compton
components~\citep{2013ApJ...771L..13T}, in good qualitative agreement
with the above arguments.

Other potential radiative signatures of the shock microphysics come
from the small-scale nature of the turbulence and its long-term
evolution in the blast. As discussed in Section~\ref{sec:PIC_mag}, one
notably expects this turbulence to relax through collisionless damping
on hundreds of $c/\omega_{\rm
  p}$~\citep{chang_08,keshet_09,2015JPlPh..8145101L} while the
electrons typically cool on much longer length scales. In GRB external blast waves, the shocked region is typically $7-9$
orders of magnitude larger than $c/\omega_{\rm p}$ in size, which
leaves room for a substantial evolution of $\epsilon_B$, even if it
decreases as a mild power-law in distance from the shock, as suggested
by the above studies. Since the electron cooling length depends on the
inverse of the electron Lorentz factor, particles of different initial
Lorentz factors emit their energy in regions of different magnetic
field strength, leading to a non-standard synchrotron
spectrum~\citep{rossi_rees_03,2007Ap&SS.309..157D,lemoine_12}, which
could in principle be used as a tomograph of the evolution of the
micro-turbulence downstream of the shock. Interestingly,  in this picture the decay
index of the turbulence is related to the
long-wavelength content of the power spectrum of magnetic fluctuations
at the shock front, which is unknown so far, as it is known to be
modified by the acceleration of higher energy
particles~\citep{keshet_09}. Finally, it is interesting to note that
the recent broad-band analysis of GRB afterglows seen from
the radio up to GeV energies has indeed revealed spectral signatures
of a decaying magnetic field~\citep{2013MNRAS.435.3009L}, with a decay
law  scaling with distance from the shock roughly as
$\Delta x^{-0.5}$ ($\Delta x$ being the proper distance to the shock in the downstream frame).

As discussed in Section~\ref{sec:PIC_mag}, there are alternative
possibilities however; it has been suggested for instance that the
turbulence could evolve in a self-similar way as a function of
distance to the shock, maintaining a uniform $\epsilon_B$ thanks to an
inverse cascade process~\citep{2007ApJ...655..375K}. 
It
is also possible that external sources seed the blast with a large
scale long-lived turbulence, \eg through a Rayleigh-Taylor instability
at the contact discontinuity~\citep{2009ApJ...705L.213L} or through
small scale dynamos following the interaction of the shock front with
external
inhomogeneities~\citep{sironi_goodman_07,couch_08}. Hopefully, future
high accuracy observational data will provide diagnostics which can be
confronted with numerical simulations.

The possibility that the small scale nature of the turbulence gives
rise to diffusive (or jitter) synchrotron radiation rather than
conventional synchrotron radiation has also attracted
attention~\cite[\eg][]{medvedev_00,medvedev_06,fleishman_06a,2011ApJ...731...26M,2011ApJ...737...55M,2013ApJ...774...61K}. In
particular, jitter radiation has been proposed as a solution for the
fact that GRBs prompt spectra below the peak frequency are
not always compatible with the predictions of synchrotron emission
(the so-called ``line of death'' puzzle, see \citet{preece_98}). In
the jitter regime, particles are deflected by less than $1/\gamma$
($\gamma$ is the electron Lorentz factor) as they cross a wavelength
$\lambda_{\delta B}$, implying that coherence of the emission is
maintained over several coherence cells of the turbulence. This regime
thus takes place whenever the wiggler parameter $a\,\equiv\, e\delta
B\lambda_{\delta B}/mc^2\,\ll\,1$, while the standard synchrotron
approximation becomes valid in the opposite limit. However, it is easy
to verify that in the vicinity of the shock
$a\,\sim\,\overline\gamma_{\vert \rm sh}$, with
$\overline\gamma_{\vert\rm sh}$ the average Lorentz factor of the
supra-thermal electrons in the shock rest frame, suggesting that
jitter signatures must be weak.

\begin{figure}[!tbp]
\begin{center}
\PNGfigure{\includegraphics[width=0.9\textwidth]{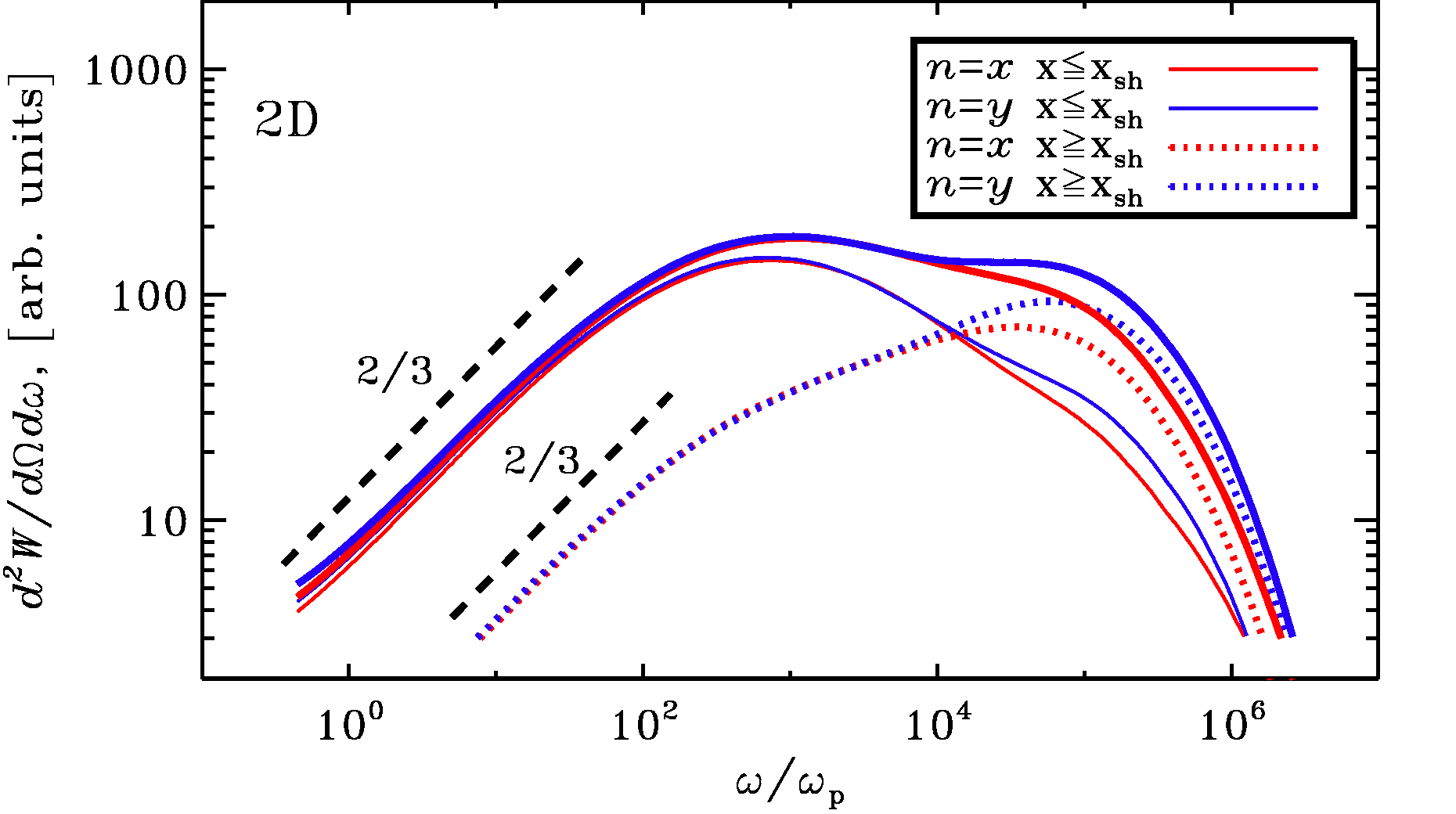}}
\caption{\footnotesize{\textit{Ab initio} photon spectrum (thick solid
    lines) from the 2D PIC simulation of an unmagnetized (\ie
    $\sigma=0$) pair  shock. Red lines are for head-on emission ($\hat{n}=\hat{x}$, along the
    shock direction of propagation), blue lines for edge-on emission
    ($\hat{n}=\hat{y}$, along the shock front). The slope at low frequencies is $2/3$
    (black long-dashed lines), proving that the spectra are consistent
    with synchrotron radiation from a 2D particle distribution (in 3D,
    the predicted slope of 1/3 is obtained).  By separating the
    relative contribution of downstream ($x\leq x_{\rm sh}$; thin solid lines) and
    upstream ($x\geq x_{\rm sh}$; dotted lines) particles, one sees that upstream
    particles contribute significantly to the total emission (thick
    solid lines), especially at high frequencies. Frequencies are in
    units of the plasma frequency $\omega_{\rm p}$.}}
\label{fig:radiation1}
\end{center}
\end{figure}

The absence of jitter radiation in relativistic shocks has been demonstrated from first principles by
computing the radiation from particles in PIC
simulations~\citep{sironi_spitkovsky_09b}, which produce spectra
entirely consistent with synchrotron radiation in the fields generated
by the Weibel instability (Fig.~\ref{fig:radiation1}). The so-called
``jitter'' regime is recovered only by artificially reducing the
strength of the fields, such that the parameter $a$ becomes much
smaller than unity. So, if the GRB prompt emission results
from relativistic unmagnetized shocks, it seems that resorting to the
jitter regime is not a viable solution for the ``line of death''
puzzle. At frequencies above the peak, the synthetic spectra from PIC
simulations show, somewhat unexpectedly, that the contribution of the
upstream medium to the total emission is not negligible
(Fig.~\ref{fig:radiation1}), yet it is omitted in most models. This
causes the radiation spectrum to be flatter than the corresponding
downstream spectrum, thus partly masking the contribution of
downstream thermal particles.

\subsection{Radiative Signatures of Pulsar Wind Nebulae}\label{sect:pwn}
The spectrum of PWNe consists of two components, where the low energy component, most likely dominated by synchrotron, shows a cutoff at a few tens of MeV. The fact that synchrotron emission reaches these energies, despite the rapid synchrotron cooling, implies that particle acceleration in the nebula is an extremely fast process \citep{dejager_harding_92}, which challenges our understanding of particle acceleration in relativistic shocks.

Around the equatorial plane of obliquely-rotating pulsars, the wind consists of toroidal stripes of opposite magnetic polarity, separated by current sheets of hot plasma. It is still a subject of active research whether the alternating stripes will dissipate their energy into particle heat ahead of the termination shock, or whether the wind remains dominated by Poynting flux till the termination shock \citep[][]{lyubarsky_kirk_01,kirk_sk_03,sironi_spitkovsky_11b}. If the stripes are dissipated far ahead of the termination shock, the upstream flow is weakly magnetized and the pulsar wind reaches a terminal Lorentz factor (in the frame of the nebula)
$
\gamma_r\sim L_{sd}/m_e c^2 \dot{N}\simeq3.7\times 10^{4} L_{sd,38.5}\dot{N}_{40}^{-1}~,
$
where $L_{sd}\equiv 3 \times 10^{38}L_{sd,38.5}\unit{erg s\,s^{-1}}$ is the spin-down luminosity of the Crab (the Crab Nebula is the prototype of PWNe), and $\dot{N}=10^{40}\dot{N}_{40}\unit{s^{-1}}$ is the particle flux entering the nebula, including the radio-emitting electrons \citep{bucciantini_11}.

For electron-positron flows, as appropriate for pulsar winds, the maximum particle Lorentz factor in the downstream frame increases with time as $\gamma_{max}\sim 0.5 \,\gamma_r\, (\omega_{\rm p} t)^{1/2}$ (see Section \ref{PIC}). The plasma frequency $\omega_{\rm p}$ can be computed from the number density ahead of the termination shock, which is $n_{{\rm TS}}=\dot{N}/(4 \pi R_{\rm TS}^2 c)$, assuming an isotropic particle flux. Here, $R_{\rm TS}\equiv3\times10^{17}R_{\rm TS,17.5}\unit{cm}$ is the termination shock radius. Balancing the acceleration rate with the synchrotron cooling rate in the self-generated Weibel fields, the maximum electron Lorentz factor is
\be
\gamma_{sync,e}\simeq3.5\times10^{8}L_{sd,38.5}^{1/6}\dot{N}_{40}^{-1/3} \epsilon_{B,-2.5}^{-1/3}R_{\rm TS,17.5}^{1/3}~.
\ee
A stronger constraint comes from the requirement that the diffusion length of the highest energy electrons be smaller than the termination shock radius (\ie a confinement constraint). Alternatively, the acceleration time should be shorter than $R_{\rm TS}/c$, which yields the critical limit
\be
\gamma_{\mathit{conf,e}}\simeq1.9\times10^{7}L_{sd,38.5}^{3/4}\dot{N}_{40}^{-1/2}~,
\ee
which is generally more constraining than the cooling-limited Lorentz factor $\gamma_{sync,e}$.
The corresponding synchrotron photons will have energies
\be
\!\!\!h \nu_{\mathit{conf,e}}&\simeq&0.17\,L_{sd,38.5}^{2}\dot{N}_{40}^{-1}\epsilon_{B,-2.5}^{1/2}R_{\rm TS,17.5}^{-1}\unit{keV}
\ee
 which are apparently too small to explain the X-ray spectrum of the Crab, extending to energies beyond a few tens of MeV.
 We conclude that Fermi acceleration at the  termination shock of PWNe is not a likely candidate for producing X-ray photons via the synchrotron process, and valid alternatives should be investigated.

One possibility -- magnetic dissipation of the striped pulsar wind in and around the shock front itself -- has been extensively studied, with the conclusion that particle acceleration along extended X-lines formed by tearing of the current sheets may contribute to the flat particle distribution (with spectral index $s_\gamma\simeq1.5$) required to explain the far infrared and radio  spectra of PWNe \citep[\eg,][]{lyubarsky_03, sironi_spitkovsky_11b,sironi_spitkovsky_12}. Indeed, hard particle spectra are found to be a generic by-product of magnetic reconnection in the relativistic regime appropriate for pulsar winds \citep[][see also Kagan et al. (2015) in the present volume]{sironi_spitkovsky_14,sironi_15}. However, further acceleration to gamma-ray emitting energies by the Fermi process cannot occur in the shock that terminates the pulsar wind, if particle scattering depends only on the self-generated turbulence.

Yet, the steady-state  hard X-ray and gamma-ray spectra of PWNe do look like the consequences of Fermi acceleration -- particle distributions with $s_\gamma \simeq 2.4$ are implied by the observations. In this regard, we argue that the wind termination shock might form in a macroscopically turbulent medium, with the outer scale of the turbulence driven by the large-scale shear flows in the nebula \citep{komissarov_04,delzanna_04,camus_09}. If the large-scale motions drive a turbulent cascade to shorter wavelengths, back-scattering of the particles in this downstream turbulence, along with upstream reflection by the transverse magnetic field of the wind, might sustain Fermi acceleration to higher energies.

Another ``external'' influence of reconnection on the shock structure, that might lead to particle acceleration to higher energies, may be connected to the accelerator behind the recently discovered gamma-ray flares in the Crab Nebula \citep{abdo_11}.
Runaway acceleration of electrons and positrons at reconnection X-lines, a linear accelerator, may inject energetic beams into the shock, with the mean energy per particle approaching the whole open field line voltage, $\gtrsim 10^{16}\unit{V}$ in the Crab \citep{arons_12}, as required to explain the Crab GeV flares.  This high-energy population can drive cyclotron turbulence when gyrating in the shock-compressed fields, and resonant absorption of the cyclotron harmonics can accelerate the electron-positron pairs in a broad spectrum, with maximum energy again comparable to the whole open field line voltage \citep{hoshino_92,amato_arons_06}.

\section{Conclusions}\label{conc}
There has been significant progress in our understanding of relativistic shocks in recent years, thanks to both analytical work 
and numerical simulations. The highly nonlinear problem of particle acceleration and magnetic field generation in shocks --- with the accelerated particles generating the turbulence that in turn mediates their acceleration --- is being tackled from first principles, assessing the parameter regime where particle acceleration in relativistic shocks is efficient. In this chapter, we have described the basic analytical formalism of test particle acceleration in relativistic shocks, leading to the ``universal'' energy slope $s_\gamma\simeq 2.2$ in the ultra-relativistic limit;
we have unveiled the most relevant plasma instabilities that mediate injection and acceleration in relativistic shocks; and we have summarized recent results of large-scale PIC simulations concerning the efficiency and rate of particle acceleration in relativistic shocks, and the long-term evolution of the self-generated magnetic turbulence. Our novel understanding of particle acceleration and magnetic field generation in relativistic shocks has profound implications for the modeling of relativistic astrophysical sources, most importantly PWNe, GRBs, and AGN jets.

\vspace{0.3in}
{\bf Acknowledgments:} We gratefully thank Boaz Katz, Guy Pelletier, Anatoly Spitkovsky and Eli Waxman for their collaboration on many of the issues discussed here. U.K. is supported by the European Union Seventh Framework Programme (FP7/2007-2013) under grant agreement n\textordmasculine ~293975, by an IAEC-UPBC joint research foundation grant, and by an ISF-UGC grant. M.L. acknowledges support by the ANR-14-CE33-0019 MACH project.

\bibliographystyle{aps-nameyear}      



\end{document}